\documentclass[aps,prl,preprint,groupedaddress]{revtex4}
\def\ll{_}
\def\uu{^}
\def\z{\zeta}
\def\g{\gamma}

\def\d{\delta}

\newcommand{\heading}[1]{\begin{center}\it {#1} \rm \end{center}}
\def\sun{\odot}

\newcommand\ba{\begin{eqnarray}}
\newcommand\ea{\end{eqnarray}}
\newcommand{\bbb}{\ba\begin{array}{c}}
\newcommand{\eee}{\nonumber\end{array}\ea}
\newcommand{\een}[1]{\label{#1}\end{array}\ea}
\def\hh{{\frac 1 2}}

\def\t{\tau}

\usepackage{amsmath,amssymb}
\usepackage{epsfig,psfrag}
\usepackage{color}
\definecolor{refcol}{rgb}{0,0.7,0.4}
\definecolor{eqcol}{rgb}{.6,0,0}
\definecolor{purple}{cmyk}{0,1,0,0}
\usepackage[colorlinks=true]{hyperref}

\makeatletter
\gdef\@citecolor{refcol}
\gdef\@linkcolor{eqcol}
\def\colorlinkspurple{\gdef\@urlcolor{purple}}
\def\colorlinksblue{\gdef\@urlcolor{blue}}
\def\colorlinksred{\gdef\@urlcolor{red}}
\makeatother

\def\revise#1       {\raisebox{-0em}{\rule{3pt}{1em}}% 
                     \marginpar{\raisebox{.5em}{\vrule width3pt\ 
                     \vrule width0pt height 0pt depth0.5em 
                     \hbox to 0cm{\hspace{0cm}{% 
                     \parbox[t]{4em}{\raggedright\footnotesize{#1}}}\hss}}}}

\begin{document}

\preprint{hep-th/0508161}

\title{Dark Matter and The Anthropic Principle}


\author{Simeon Hellerman}
%\thanks{}
\affiliation{School of Natural Sciences, Institute for Advanced 
Study, Princeton, New Jersey, USA}
\author{Johannes Walcher}
\affiliation{School of Natural Sciences, Institute for Advanced 
Study, Princeton, New Jersey, USA}


\date{August 22, 2005}

\begin{abstract}
We evaluate the problem of galaxy formation in the landscape 
approach to phenomenology of the axion sector.  With other parameters of
standard $\Lambda$CDM cosmology held fixed,
the density of cold dark matter is bounded below relative 
to the density of baryonic matter by the requirement that structure
should form before the era of cosmological constant domination of the
universe. 
Galaxies comparable to the Milky Way can only form if the ratio
also satisfies an upper bound.
The resulting constraint
on the density of dark matter is
too loose to 
select a low axion decay constant or small initial displacement
angle on anthropic grounds.
\end{abstract}

\pacs{11.25.-w, 14.80.Mz, 95.35.+d, 98.80.Bp}
\keywords{Axions, Dark Matter, Anthropic Principle, String cosmology}

\maketitle

%\subsection{}
%\subsubsection{}

\section{Introduction}

In the ``landscape'' approach to string phenomenology \cite{landscape0} - 
\cite{landscape3}, one starts with an assumption of a large number, $\sim 
10^{\text{several hundred}}$, of discrete vacua of the fundamental theory, a 
picture which is supported by several estimates based on counting solutions 
to the effective action of string theory on Calabi-Yau manifolds. The goal
of the program is to explain low energy phenomena not as a unique and direct
consequence of Planck scale physics, but as one realization 
among a rich set of possibilities. An important input in this approach 
is the ``anthropic cut'': The only potentially phenomenologically relevant 
parts of the landscape are the vacua whose effective physics allows 
for the occurrence of biologically complex life. 

One fruitful application of the anthropic approach has been Weinberg's 
reasoning \cite{weinberg} leading to a bound on the cosmological 
constant from the minimal condition that galaxies must have 
enough time to form. From this apparent success we can derive a general 
lesson about the application of anthropic reasoning: the simplest anthropic 
constraints are those related to physics which is almost decoupled 
from phenomena at ordinary scales.  Attempts to extract anthropic
predictions for other (particle or cosmological) standard model 
parameters is hampered by our ignorance of the distribution of
vacua, but also by the intricacy and interconnectedness of
the effects of those parameters on the development of life,
and on each other at different scales via RG evolution.

One highly decoupled sector where anthropic reasoning might usefully be 
applied is that of the Peccei-Quinn axion, $a$. 
Postulated as a mechanism for solving the strong CP problem, 
the axion is a pseudoscalar field with an approximate shift symmetry
$a\to a + \theta$ which is broken only by the coupling 
$\Delta\mathcal L= a n{\rm tr} (F\wedge F)_{SU(3)}$. If $a$ is
dimensionless, its kinetic term, $\Delta\mathcal L_{\rm kinetic}
= f^2 (\partial a)^2$ contains a dimensionful parameter $f$, 
known as the axion decay constant. QCD instantons then induce an axion 
potential of the form
$$
V(a) = c \Lambda_{\rm QCD}^4 \sin^2(n a)
$$
where $c$ is a numerical coefficient which can be set to $1$ by
a choice of definition of the dynamical scale $\Lambda_{\rm QCD}$.
An integer rescaling of $f$ can also set $n$ to $1$. So the physical 
mass of the axion is given by by $\Lambda_{\rm QCD}^2/f$ at tree level. 
The massive parameter $f$ is determined by physics at scales
above the standard model. In a generic high-scale model,
such as any GUT or string model, the expectation would be
that $f$ should have its magnitude set by the scale at which
new physics enters---say $10^{16}$ GeV or so. This expectation
is borne out specifically in superstring models such as
the heterotic string
\cite{choiandkim}, as well as in many of the more
recently studied perturbative superstring vacua \cite{ed}.
\footnote{It was noted in 
ref. \cite{ed} that smaller values of $f$ might also be realized in
certain string models. The distribution of axion decay constants 
in, {\it e.g.}, the string landscape is not known.}

As has been realized for a long time \cite{long}, such a value of $f$ 
appears in conflict with the cosmological standard model, as 
relic axions produced from initial vacuum displacement in the 
early universe make a contribution to the dark matter density 
that exceeds the observed value by orders of magnitude.  Linde has
pioneered the approach \cite{linde} of using anthropic ideas
to loosen this bound in inflationary 
cosmology, where the homogeneous initial misalignment angle,
$\theta_0$, of the axion is a free parameter, putatively 
``environmental'', in the sense of varying from region to
region.  If $f$ is fixed at its natural value, however, inflationary 
fluctuations are too large to make this work \cite{tuwi}, a conclusion 
which possibly can be avoided in models of hybrid inflation 
\cite{linde2}. More recently, Banks and Dine \cite{badi}, and Banks, 
Dine, and Graesser \cite{badigr} have emphasized that the cosmological 
axion problem is dominated in the supersymmetric context by the Saxion 
(and other moduli), and that a satisfactory resolution might require 
a much more drastic modification of the history of the universe between 
inflation and decoupling.

While this debate is by no means settled, it shows that if the 
strong CP problem is solved by a Peccei-Quinn axion in our universe, 
realizing observational results on dark matter density will most 
likely require some degree of fine adjustments of parameters and/or 
initial conditions. In the absence of a mechanism, but in the
background of the landscape, we may ask if such adjustments can perhaps 
be justified anthropically, as has been done in \cite{linde} and 
elsewhere (see {\it e.g.}, \cite{wilczek}). While not as severe a case 
as the cosmological constant, it seems a sensible sector to apply 
anthropism, since dark matter is essentially decoupled from everyday 
low-energy physics just as dark energy is.

We will here evaluate the extent to which the requirement that 
habitable structures form bounds the ratio of dark matter to baryonic
matter in the universe. It is well accepted that in order for galaxies 
and similar astrophysical objects to form out of the primordial density 
perturbations seeded during inflation, the matter density $\rho_{\rm matter}$ 
should contain a predominant dark matter component, $\rho_{\rm DM}$, which 
drives the growth of structure between equality and decoupling. It is
also clear that without any baryons, $\rho_{b}\to 0$,
all structure would remain dark and uninhabited. A combination that works 
well is when the ratio $\zeta=\rho_{\rm DM}/\rho_{b}\approx
5$ as in our universe. It thus being clear that a universe with
$\zeta=5$ is habitable, and a universe with $\z^{-1}=0$ is
not, one wonders what range of values of $\zeta$ life can actually 
tolerate. Our chief interest here is to understand what general form
such an anthropic bound may take and where in the allowed region our
universe is situated, as well as what detailed astrophysics the
tightness of the bound depends on.

To keep control, we will fix all other cosmological parameters, such as 
baryon to photon ratio which we call $\eta \equiv n\ll b / n\ll \g$, 
scale and spectrum of initial perturbations, {\it etc.}, to the values
we have observed today. Sometimes, it will be convenient to keep the 
cosmological constant term $\rho_\Lambda$ as a free parameter in the 
discussion. As in \cite{weinberg}, it is $\rho_{\Lambda}$ which 
threatens life by halting the ultimate global formation of structure 
at a later stage in the evolution of the universe.

\section{An Elementary Bound}

\begin{figure}
\psfrag{L}{$\rho_\Lambda/M_{\rm Pl}^4$}
\psfrag{z}{$\zeta$}
\psfrag{1}{$1$}
\psfrag{11}{$10$}
\psfrag{12}{$10^2$}
\psfrag{13}{$10^3$}
\psfrag{14}{$10^4$}
\psfrag{15}{$10^5$}
\psfrag{20}{$10^{-126}$}
\psfrag{21}{$10^{-125}$}
\psfrag{22}{$10^{-124}$}
\psfrag{23}{$10^{-123}$}
\psfrag{24}{$10^{-122}$}
\psfrag{25}{$10^{-121}$}
\psfrag{27}{$10^{-120}$}
\psfrag{28}{$10^{-119}$}
\epsfig{file=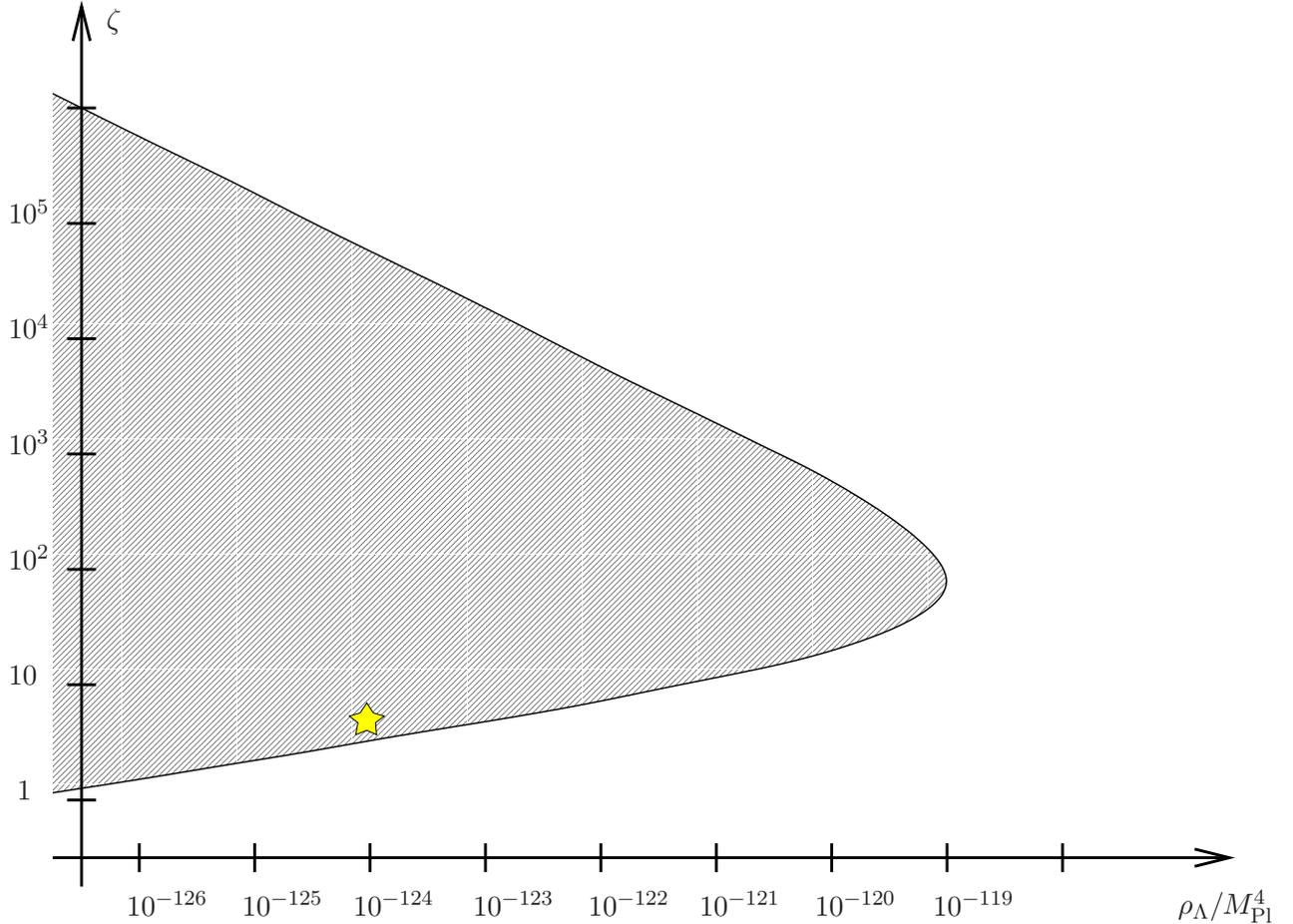,width=\columnwidth}%
\caption{\label{region} A schematic picture of the region in 
$\Lambda$-$\zeta$ space for which gravitationally bound structures 
containing $10^{11}M_{\odot}$ of baryons form. The star represents 
our universe.}
\end{figure}

At equality of matter and radiation, $\rho_{\gamma}=\rho_{\rm 
matter}=\rho_{b}+\rho_{\rm DM}\approx\rho_{\rm DM}=
\zeta\rho_{b}$. Using $\rho_{\gamma}=T_{\rm eq}^4
=T_{\rm eq} n_\gamma$, and $\rho_{b}=\mu n_{b}$, where $\mu=1$GeV
is the mass of a baryon, this gives for the temperature at equality
\footnote{We are here assuming that baryons are non-relativistic
at equality. For $T_{\rm eq}>\mu$, the matter to photon ratio
$\xi=\rho_{\rm matter}/n_\gamma=T_{\rm eq}$ is a more useful parameter
\cite{tere}, while $\zeta=\xi/(\eta \mu)$ obtains significance only once 
the temperature drops below $\mu$. The present parameterization, which 
is sensible up to $\zeta=\eta^{-1} \approx 10^9$, is more convenient 
for our later considerations. In order to avoid interfering with
big bang nucleosynthesis, one might also wish to keep $T_{\rm eq}$
below $1$MeV.}
\begin{equation}
T_{\rm eq} = \mu\eta\zeta
\end{equation}
The density perturbations, which we assume to be of inflationary origin,
can be divided roughly into two classes, depending on their mass
scale, $M$. Since perturbations only grow logarithmically in the radiation
dominated era, all density perturbations whose physical size is
smaller than the horizon size at equality have their primordial strength 
$\delta\approx\delta_0 = 10^{-5}$. Perturbations which are superhorizon
at equality will reach their scale-invariant amplitude when they enter the
horizon and can be ascribed a strength $\delta \approx \delta_0 (\lambda/
H_{\rm eq}^{-1})^{-2}$ at equality. Here, $\lambda= (M/\rho_{\rm eq})^{1/3}$ 
gives the relation between the size, $\lambda$, of a perturbation and its 
mass scale, $\rho_{\rm eq}=T_{\rm eq}^4 = (\mu\eta\zeta)^4$ is the energy 
density at equality, and $H_{\rm eq}^{-1}=(G\rho_{\rm eq})^{-1/2}$ is the 
horizon size. Thus,
\begin{equation}
\delta_{M,{\rm eq}} \approx
\begin{cases}
\delta_0 \qquad &\lambda_{M,{\rm eq}} < H^{-1}_{\rm eq}\\
\delta_0 \bigl(M^{1/3}(\mu\eta\zeta)^{2/3}/M_{\rm Pl}\bigr)^{-2}  
\qquad  &\lambda_{M, {\rm eq}} > H^{-1}_{\rm eq}
\end{cases}
\end{equation}
In the matter dominated era, the strength of the perturbations grows 
linearly with the scale factor of the universe, $\delta\propto a$. The
non-linear regime is reached when $a/a_{\rm eq}\approx 1/\delta_{M,{\rm eq}}$, 
after which the structure breaks away from the overall expansion of the 
universe. The celebrated Weinberg bound expresses the fact that this should 
happen before the universe is dominated by vacuum energy, $\rho_\Lambda$,
lest acceleration disrupt the forming structure. So, structures of scale
$M$ have time to form between equality and cosmological constant domination
if and only if
\bbb
\rho_\Lambda \lesssim \rho_{\rm eq} \bigl(\delta_{M,{\rm eq}}\bigr)^3
\een{linbound}
Now, we have to decide what scale of structure is required for life,
and how this depends on $\zeta$. Beyond its usual murkiness, this
question is even more delicate in the universes that we are envisaging, 
because the structures may look quite different from those that form
with our value of $\zeta$, as we will describe in more detail below.
As an example, we can consider the fate of a perturbation that
has a chance of evolving to a galaxy like ours. This will give us
our strongest bound on $\zeta$, and useful expectations for a more
careful study (see Fig.\ \ref{region}).

The Milky Way contains about $M_{\rm gal}= 10^{11} M_{\odot}$ worth of 
baryons and so corresponds to a total mass scale $M = \zeta M_{\rm gal}$. 
Inserting this into (\ref{linbound}) implies the bounds,
\begin{equation}
\rho_\Lambda \lesssim
\begin{cases}
(\mu \eta)^4 \delta_0^{3} \;\zeta^4 \qquad & 
\zeta M_{\rm gal}^{1/3}(\mu\eta)^{2/3} / M_{\rm Pl} < 1 \\
M_{\rm Pl}^6 M_{\rm gal}^{-2} \delta_0^3\;\zeta^{-2}
\qquad & \zeta M_{\rm gal}^{1/3}(\mu\eta)^{2/3} / M_{\rm Pl} > 1
\end{cases}
\label{bound}
\end{equation}
In other words, if the perturbation giving rise to our galaxy is subhorizon
at equality, it enters the non-linear regime after the energy density has
dropped by a fixed amount. Since the energy density at equality scales with
the fourth power of $\zeta$, the cosmological constant can be correspondingly
larger. If our galaxy is superhorizon size at equality, the strength of
the corresponding perturbation is down by a factor of $\zeta^{-2}$, and
it takes correspondingly longer to grow to non-linearity. The cosmological
constant cannot be too large.

Besides its simplicity, the interest of this derivation is that, for 
fixed $\Lambda$, it yields both a lower and an upper bound on $\zeta$. 
Numerically, in our universe, $\eta=10^{-9}$, $\zeta=5$, $\delta_0= 
10^{-5}$, $\lambda_{\rm gal}/H_{\rm eq}^{-1}\approx 5\times 10^{-2}$, while the 
vacuum energy density is comparable to its upper bound. (We are using numbers
from, {\it e.g.}, \cite{pars}.) Therefore, if we increased $\zeta$ by a factor 
of $20$, the perturbation corresponding to our galaxy would have extended 
up to the horizon at equality, and $\Lambda$ could have been $\sim 10^5$ 
times larger. Increasing $\zeta$ by another factor of $400$ brings back 
the bound on $\Lambda$ to the familiar value. Thus, for fixed $\Lambda$, 
the existence of our Milky Way can tolerate a value of $\zeta$ in the 
range $5\lesssim \zeta \lesssim 8 \times 10^4$.

The simplest version of the upper bound is to say that for fixed
$\Lambda$, the maximum mass of any structure which can form
is given by $M\uu{(max)} \lesssim M\ll{pl}\uu 3~\d\ll 0\uu{3/2} ~
\rho\ll\Lambda\uu{-\hh}$, and the largest number of baryons which can
exist in a gravitationally bound structure is
$M\ll{b}\uu{(max)} = \zeta\uu{-1}~M\uu{(max)}$.
For $\Lambda$ fixed to the observed value but
letting $\zeta$ vary, we find that $M\uu{(max)}
= 9.6 ~\times 10\uu{15} M\ll\sun$ in our universe for
the total mass, and $M\uu{(max)}\ll{b} = 1.6 \times
10\uu{15} M_\sun$ for the baryonic mass, where we have taken $\z = 5$.
% DON'T REMOVE ...
%Here we have used $1 M\ll\sun = 1 . 99 \times 10\uu{30}~
%kg = 1.12 \times 10\uu{57}~GeV$ and $M\ll{pl}
%= 2.18 \times 10\uu {-8} kg =  1.22 \times 10\uu{19}~GeV$,
%which means 
%$M\ll\sun / M\ll{pl} = 0.91 \times 10\uu{38}$.
%For the vacuum energy we use $\rho\ll \Lambda \simeq
%2.8~\times~ 10\uu{-47} ~GeV\uu 4$.  We also
%take $\d\ll 0 = 10\uu{-5}$ so $\d\ll 0\uu{3/2} = 3.16 \times
%10\uu{-8}$.
% ... THESE COMMENTS
Fixing $\rho\ll\Lambda$ and $\d\ll 0$ to what we observe and demanding 
$10\uu{11} M\ll\sun$ as a minimum, we recover the conclusion that we 
could tolerate a value of $\zeta$ as large as $ M\ll{pl}\uu 3 \d\ll 
0 \uu{3/2} \rho\ll\Lambda\uu{-\hh} / (10\uu{11} M\ll\sun) = 9.4 \times 
10\uu 4\approx 10^5$.

%\begin{widetext}
%\end{widetext}

\section{Refinements}

\heading{Nonlinear analysis of structure formation}

In this derivation, we have used a linear analysis of the
growth of density perturbations into gravitationally
bound structures.  The motivation for this heuristic treatment
is that local self-gravity will only be important when
$\d \rho / \rho$ is of order 1; and due to the universal
attractiveness of gravity, the growth of structures can
only be hastened when the perturbation strength is large,
rather than retarded.  This suggests that the time scale
for the development of a bound structure should not be
significantly longer or shorter than the time scale for the
perturbation to reach unit strength.

We point out that a more careful linear treatment alters our 
bound only by a factor of order 1.  A formula due to Weinberg
\cite{weinberg} states that the minimum density perturbation 
which can collapse gravitationally must obey the bound
\bbb
\rho\ll \Lambda \leq \frac {500} {729} ~\rho\ll {\rm eq}
~\d\ll{M,\rm eq}\uu 3~,
\een{nonlinbound}
which gives a version of equation
(\ref{linbound}) with slightly better
numerical precision.  The effect is merely to change the
upper bound on $\zeta$ by a factor of 
 $ \left ( \frac  {500} {729}  \right )\uu{ \frac 1 4} 
\simeq 0.91
$,
or $ \left ( \frac  {500} {729}  \right )\uu{- \frac 1 2}
\simeq 1.21
 $,
respectively, in the two regimes.  As these factors are
smaller than several other numerical imprecisions in our
argument, we are justified in neglecting them before and
hereafter.

\heading{Probabilities}

In Weinberg's original derivation \cite{weinberg}, the upper bound on 
the cosmological constant turned out 2--3 orders of magnitude bigger 
than the then valid observational upper bound (which is the now 
measured value).  The conceptual input into the derivation is that
the \it earliest \rm galaxies should form before the era of 
$\Lambda$-domination, and indeed we can recover Weinberg's bound by 
recalling that galaxies at high redshift $z\approx 4$ are seeded by 
density perturbations of strength $10^{-4}$. Such density perturbations 
lie on the tail of a distribution with mean strength $10\uu{-5}$, 
as determined by precision measurements of the CMB.

One might try to improve the bound by asking whether a set of
cosmological parameters which leave more time for galaxies to form
might be in some sense more anthropically favorable than parameters
which leave just the minimal amount of time \cite{vilenkin,weinberg2}. 
A really scrupulous attempt to answer this question is hampered by 
the lack of an \it a priori \rm notion of relative anthropic
favorability of two acceptable universes.  Weinberg \it et al. \rm
\cite{weinberg3} improve the bound on $\rho\ll\Lambda$ by assigning 
universes an anthropic probability based on the number density of 
observers. Since most galaxies form from the peak of the distribution,
it is no surprise that the most likely value of $\Lambda$ is equal
to its upper bound with the amplitude of fluctuations set to their
mean value \cite{weinberg3}.

The logical foundation of such a weighting is unclear. Indeed the notion 
of number density of observers is gauge dependent, and is defined in 
\cite{weinberg3} in an \it ad hoc \rm way relative to FRW time-slices.
However one does not wish to argue with success; the approach of
\cite{weinberg3} predicts a vacuum energy equal to $o(1)$ of the 
critical density, matching the value of $\rho\ll\Lambda \simeq 0.71 
\rho\ll{crit.}$. Perhaps something valuable can be learned by putting 
this approach on firmer logical ground.

For the distribution of both parameters $(\zeta,\Lambda)$, we 
would consider it premature to do a similar computation, not least
because it is difficult to guess an appropriate a priori probability
distribution without specifying a microscopic model for dark matter.
(The cosmological constant is slightly more favorable in this regard 
because one can argue \cite{weinberg3} that any \it a priori \rm 
probability distribution should be flat in the anthropically allowed 
region, which is narrow and far away from the natural value.)
If we assume, however, that dark matter is an axion relic originating
in string theory, it seems that the peak of the a priori distribution
should lie at large values of $\zeta$, with at least a power-law
cutoff for smaller values. It appears unlikely that weighting with 
the number of observers can reverse this trend and produce a peak
close to the values we observe (the star in Fig.\ \ref{region})---%
unless one of the astrophysical effects discussed below serves to 
eliminate observers for large values of $\zeta$ altogether.

\section{Life in a baryon-poor galaxy?}

As we have seen, the requirement of gravitationally bound structures
containing a certain fixed number $N$ of baryons
imposes a sharp cutoff on the allowed values of $\zeta$, the actual 
value $\zeta(N)$ of the cutoff being $N$-dependent. 
For purposes of obtaining a first impression, we have taken
$N = 10\uu{11} ~M\ll{\sun} /\mu$. This is a very strong requirement, 
and is essentially biased by the fact that the only life we know of 
is the one on Earth. The choice ignores the possibility 
that observers could evolve in a galaxy with far fewer baryons --- 
$N\sim 10\uu 6 M\ll{\sun} / \mu$, for example.

Moreover, our discussion so far has ignored the stages of structure 
evolution which happen after the inhomogeneities break away from the 
overall expansion of the universe. The cosmological constant is 
to a good degree irrelevant after this stage because its gravitational 
pull is so weak. The dark matter to baryon ratio, however, controls the 
type and succession of structures that do manage to form, and thereby
has a significant influence on both the top-down (first stars to quasars 
to galaxies to clusters) as well as the bottom-up (galaxies to stars 
to planets) branches of the subsequent evolution.

This impact is of course very difficult to evaluate, given that the
details of non-linear formation of structure are not under complete
control even in our universe! Nevertheless, the relevant qualitative
features are reasonably well understood, and extrapolating to
extreme values of $\zeta$ will allow us to at least address some
of the issues.

\heading{Cooling}

The first thing that has to happen after the density perturbations 
have become non-linear is that the baryonic gas that has fallen
into the collapsed dark matter halos and shock-heated to the
virial temperature must be able to cool efficiently in order to
contract beyond virialization, fragment, and ignite stars after
reaching nuclear densities \cite{osre}.

It is often stated that to measure the efficiency of cooling, one 
should compare the dominant cooling rate $\tau_{\rm cool}^{-1}$ with 
the dynamical time scale $\tau_{\rm dyn}=(G\rho_{\rm vir})^{-1/2}$. 
If $\tau_{\rm cool} \gg \tau_{\rm dyn}$, cooling is considered inefficient. 
It is obvious that in our universe, structures with cooling times much 
longer than the age of the universe simply have not had the time to 
evolve until today. In some other scenarios, one can also imagine
that structures of some given size should cool significantly before 
the next bigger structures collapse onto them, since otherwise the
smaller structures would not survive as distinct entities.
In alternate universes of the type we are considering here, however,
the last structures to form before cosmological constant domination 
will not suffer from this, and it is conceivable that given sufficient 
time, they will cool and can possibly develop life. (We must, of course, 
assume that the cooling time is not competitive with the lifetime of the 
proton!)

In any event, by following the standard treatments, discussed for example 
in \cite{tere} in the context of anthropic constraints on the
amplitude of the primordial perturbations, we can estimate
how the efficiency of the cooling varies with $\zeta$. 

The dominant mechanisms which have contributed to the cooling of
the structures in our universe are atomic and molecular
line cooling of hydrogen and heavier elements as well as
bremsstrahlung resulting from collisions of constituents of the charged 
plasma in the potential well of the dark matter halo. Both 
these mechanisms depend on the density, temperature, and the ionization 
level of the gas, and hence on $\zeta$.

The virialization density $\rho_{\rm vir}$ is proportional to
the total matter density at the time of collapse, $\rho_{\rm coll}
\approx \rho_{\rm eq} \delta_{M,{\rm eq}}^3$ and the virial
temperature $T_{\rm vir} = G M \mu /r = M_{\rm Pl}^{-2} 
\mu M^{2/3} \rho_{\rm vir}^{1/3}$. As we have seen, the density
at collapse increases independent of $M$ as $\zeta^4$ until
$M\sim M_{\rm Pl}^3(\mu\eta\zeta)^{-2}$, and thereafter drops
as $M^{-2}$, independent of $\zeta$. As a result, $T_{\rm vir}
\propto M^{2/3}\zeta^{4/3}$ and $T_{\rm vir}\approx \mu\delta_0$
in the two regimes, respectively. \footnote{These estimates also imply 
that the mean density of a collapsed structure has a maximum as a function 
of the mass for fixed $\zeta$. It also has a maximum as a function of 
$\zeta$ for fixed mass. The collapse density alone therefore does not 
seem suited as anthropic gauge as in \cite{linde}.} Therefore, if we 
increase $\zeta$, the virial temperature will soon exceed $10^4K$, and 
line cooling will cease to be relevant for most structures. Cooling by 
bremsstrahlung will dominate, and we find 
\begin{equation}
\frac{\tau_{\rm brems}}{\tau_{\rm dyn}} \propto
\begin{cases}
\zeta^{-1/3} M^{1/3} \qquad &\text{$M$ subhorizon at equality}\\
\zeta M \qquad &\text{$M$ superhorizon at equality}
\end{cases}
\end{equation}
We note that for fixed mass $M$, this has a maximum as a function
of $\zeta$ with different power laws in the two regimes, very much 
as we found in (\ref{bound}). An interesting point is that for
fixed $M_{\rm gal}=M/\zeta$, the efficiency of bremscooling is
at first independent of $\zeta$, so that a baryonic structure like 
the Milky Way might indeed have a very similar cooling history in 
universes with quite different values of $\zeta$.

The main effect if we increase $\zeta$, however, will be that the
structures will soon form so early that the dominant cooling is from 
Compton scattering off the cosmic microwave background. Compton cooling 
is (in some regime) independent of the temperature and density of the 
baryons, but depends quite sensitively on the temperature of the CMB 
at the time when the structures have formed. Quantitatively, $\tau_{\rm 
comp}\propto T_{\gamma}^{-4}$, and hence
\begin{equation}
\frac{\tau_{\rm comp}}{\tau_{\rm dyn}} \propto
\begin{cases}
\zeta^{-2} \\
\zeta^{4/3} M^{5/3}
\end{cases}
\end{equation}
again with a characteristic kinked power law behavior.

In the regime of dominant Compton cooling, the effect of the CMB 
is to act as friction for the charged components in the plasma. 
In contrast to other cooling mechanisms, it is quite efficient in 
absorbing angular momentum. The baryons should therefore lose their 
angular momentum and radial kinetic energy in typical time $\t\ll{\rm 
comp}$ and slide down into the minimum of the potential
in a radially symmetric way. \footnote{Even without Compton
cooling, the presence of dark matter inhomogeneities at smaller scales 
redistributes angular momentum quite efficiently, leading to an 
``angular momentum problem'' in simulations of galaxy formation.}
At the bottom of the potential, they will have little angular momentum 
support, and fragmentation into stars is also likely to be inhibited
if the collapse is sufficiently isothermal so that the Jeans mass
does not decrease too rapidly.

It is plausible that the final state of such a collapse is one where 
the bulk of the baryonic matter forms a supermassive black hole (SMBH), 
possibly with a brief intermediate stage of life as a supermassive 
star. Indeed, it is believed that most larger galaxies in our universe 
have an SMBH at their center. As has recently become clear, for instance
in the celebrated $M$-$\sigma$ relation \cite{feme} linking the
hole mass to the velocity dispersion of the central region of the galaxy,
the history of these black holes is intrinsically linked to the formation
of the galaxy itself. The SMBH can grow by accretion or mergers but
most models assume a sizable seed black hole whose likely origin is
the collapse of gas under conditions with inhibited star formation.
(See \cite{smbh} for a short list of references.) If Compton cooling 
in addition withdraws angular momentum support, collapse to a black 
hole is a very likely outcome. 

Clearly, if increasing $\zeta$ would confine baryons into black holes, 
this would be a strong anthropic basis for selecting universes
with roughly equal proportions of baryons and dark matter. At this
stage, however, it seems that numerical simulations would be needed
to confirm whether SMBHs are a reasonable scenario. \footnote{One
of the bigger uncertainties is what fraction of baryons would initially
collapse to the SMBH. If a significant fraction remains in the halo,
it would be subject to the usual accretion and feedback processes.}

\heading{Supernova pressure}

One argument that is often cited to justify the claim that galaxies with 
fewer that $10^6 M_\sun$ of baryons are unlikely to support life is that 
in our universe, such ``galaxies'' reside inside of halos of roughly the 
same size and have a much more shallow gravitational potential. As a 
consequence, when the first stars are formed, pressure created by 
supernova explosions are powerful enough to eject gas (as well as the 
heavy elements produced in the supernov\ae, which are plausibly necessary 
for life dependent on an interesting chemistry) from the galaxy, thus 
reducing the prospects of forming a second generation of stars, with 
planets around them. 

A similar mechanism would probably provide a lower cutoff on the mass 
of baryon-containing galaxies for larger values of $\z$, but the formation 
of initial baryonic structure is not under good analytic control, so we 
do not know how to compute the dependence of this effect on $\z$. The 
all-important 
\colorlinkspurple 
\href{http://www.mpa-garching.mpg.de/~vdbosch/RTNfeedback.html}
{feedback processes} are also likely to differ. 
\colorlinksred
This would be another good direction for future study in the subject of 
anthropic constraints on dark matter.
\footnote{A second effect that is believed to inhibit efficient formation 
of galaxies with baryonic mass below $10^6M_\sun$ is that smaller gas 
clouds can not shield themselves from the radiation that reionizes the
universe at a redshift $z\approx 6$. As a consequence, atomic and molecular 
line cooling will be much less efficient and the galaxies would take too 
long to collapse beyond virialization. As we have explained, cooling is a 
very sensitive issue which depends on the succession of structure formation. 
With different values of $\zeta$, baryon-poor galaxies might still be able to
cool efficiently, even after the inter-galactic medium has been reionized.}

\section{Conclusions}

In this paper we have described the likely evolution of universes with values 
of $\z$ greater than in our own universe, with other parameters of $\Lambda$CDM 
cosmology held fixed.  Moderately larger values of $\z$ allow the formation 
of structure with astrophysical conditions similar to those in our own 
galaxy.

Some uncertainties remain. Anthropic constraints on the ratio of dark matter 
to baryonic matter appear too weak to force $\z$ as low as we observe it.  
Lacking a detailed understanding of the evolution of baryon-poor dark matter 
halos, one can impose looser or more stringent assumptions on the conditions 
necessary for life; however no reasonable assumption appears stringent enough 
to force the upper bound on $\z$ lower than $\sim 10\uu 5$.

On the less restrictive end, one can explore the possibility
that baryonic structures with arbitrarily low mass---say,
$10\uu 6 M\ll \sun$ of baryons in a $10\uu{12} M\ll \sun$ halo---%
can ultimately cool and ignite stars, though perhaps on
a time scale far longer than the age of our current universe.
One relevant question is the ability of such low proportions of
baryons to form gravitationally bound structures inside the halo
in the first place, whether or not they can eventually cool and form
stars.

This question has recently been answered in the affirmative
by the discovery of a ``dark galaxy'' in the Virgo cluster
\cite{dark}
---a gravitationally bound structure of $4 \times 10\uu 7 M\ll \sun$
inside of a $2\times 10\uu{10} M\ll \sun$ dark matter halo.  This
galaxy, known as VIRGOHI 21, has a density and temperature
too low to cool efficiently by the available mechanism of hydrogen-line
cooling, but nonetheless the baryons have been able to separate themselves
from the ambient halo enough to form a disk whose structure is determined by
its own gravity.

On the more restrictive end, we can see that even if we require baryonic 
structures of $10^{11}M_\sun$, there is little which can impede
their formation for values of $\zeta$ up to $10\uu 5$.
Certainly we can be confident that halos containing 
the requisite number of baryons will form in this range. The 
dynamics of baryonic structure formation inside the halo
at higher values of $\z$ are not entirely clear, but
the inhomogeneities in the baryonic density can only
increase with time, and the result of these growing inhomogeneities
will likely be hydrogen-fusing stars.

Finally, we discuss the implications for the axion sector.
Vilenkin's ``mediocrity principle'' \cite{vilenkin} can be interpreted
as saying that when a parameter has a range of values which would
allow life to exist, we should expect the parameter to lie
at the point within that range which is most favored by
conventional notions of naturalness, or perhaps by
the statistics of discrete vacua in a fundamental theory.
Given our anthropic range for $\z$, what does Vilenkin's principle
tell us about axion physics?

In a model in which all dark matter is axionic,
$\zeta\propto f^{3/2} \theta_0^2$ (see, {\it e.g.}, \cite{sean}),
so a bound on $\zeta$ can be interpreted either as a bound
on the axion decay constant or on the initial displacement
angle.  Conventional naturalness and statistical arguments
would both seem to favor large values of $f$. \footnote{It
has been pointed out \cite{ed} that all known weakly coupled regions
of string theory have values of $f$ of at least
$10\uu{16}$GeV, which  leads us to suspect a statistical enhancement
at large values.}  Likewise, the statistics of initial values of the axion
push $\theta\ll 0$ towards values of $o(1)$.

The key point is that the pressures of mediocrity
on $f$ and $\theta\ll 0$ should both push $\z$ towards
the higher end of its anthropic window, which
is apparently high enough to falsify an anthropic
explanation for the observed size of $\z$.
Values of $\z$ up to $10\uu 5$ are
compatible with the evolution of
habitable stars and galaxies,
even with the conservative assumption that a galaxy
needs $10\uu{11} M\ll \sun$ of baryons in
order to support life.
The observed value $\z\sim 5$ is some $20,000$ times lower
than the upper end of the anthropic window,
meaning that neither a low value of $f$ 
nor a natural value of $f$ with a low value of $\theta\ll 0$ would make 
sense anthropically.

How firm is this conclusion?
The axion sector is sufficiently decoupled from
the standard model that we can map out with confidence
much of the cosmological history of universes with large
amounts of axionic dark matter.  There is
nothing obvious
in these alternative universes to obstruct the development
of life, implying that a non-anthropic explanation of
the smallness of $f$ and/or $\theta\ll 0$ is required.

Some aspects of this argument could be tightened.  If the
baryons in the high-$\z$ universes
manage to collapse in a sufficiently
isotropic way to proceed directly to a SMBH, this
might lower the anthropic upper limit on $\z$ to below
$10\uu 5$.  Nor can we yet estimate reliably the time scale
for baryons to cool and form stars inside halos for large
values of $\z$, though the only clearly anthropic time ceiling on 
this process would be the proton lifetime.  Some of these questions 
could be answered by numerical simulations,
similar to those by which we have learned about
structure formation in our own universe.

%\appendix
%\section{}

\begin{acknowledgments}
\smallskip
{\bf Acknowledgments.} 
We would like to thank Bobby Acharya, Michael Dine, Matt Kleban, Scott Thomas, 
Ra\'ul Rabad\'an, Nathan Seiberg, Peter Svr\v cek and Edward Witten for 
valuable discussions, and Jakob Walcher for important input and comments
on the manuscript.  We would also like to thank the astrophysics 
groups at the Institute for Advanced Study and Princeton University for
extremely helpful comments, 
especially Neal Dalal, Jeremy Goodman, Andrew M$\uu{\rm{\underline 
{c}}}$Fadyen, Carlos Pe\~na-Garay, Enrico Ramirez-Ruiz, Aristotle Socrates 
and Dmitri Uzdensky. This work was supported in part by 
the DOE under grant number DE-FG02-90ER40542.
\end{acknowledgments}

\end{document}